\documentclass[amsmath,amssymb,aps,twocolumn,prb]{revtex4-1}

\usepackage[usenames]{xcolor}
\usepackage[colorlinks=true,citecolor=blue,linkcolor=magenta]{hyperref}

\usepackage{titlesec}
\definecolor{natureblue}{HTML}{343796}
\titleformat{\section}[display]{\vspace{-1em}}{}{0pt}{\normalfont\sffamily\color{natureblue}}[\vspace{-3pt}\hrule]
\usepackage{setspace}
\makeatletter
\def\frontmatter@abstractfont{\sffamily\color{natureblue}\setstretch{1.5}}%
\def\frontmatter@title@format{\noindent\huge\sffamily}{}%
\def\frontmatter@authorformat{\vspace{1em}\noindent\color{natureblue}\Large\sffamily}%
\def\frontmatter@affiliationfont{\vspace{1em}\color{black}\noindent\normalsize\sffamily}%
\def\frontmatter@above@affiliation@script{\vspace{1em}\noindent}%
\def\frontmatter@makefnmark{}
\renewcommand*\frontmatter@date[2][\Dated@name]{\def\@date{}}%
\makeatother

\usepackage{amsmath,amsfonts,amssymb,amsthm,graphics,graphicx,epsfig,times,bbm}

\renewcommand{\i}{\mathrm{i}} 
\newcommand{\tr}{\text{tr}}

\definecolor{jens}{rgb}{1,0,0}
 
\definecolor{mathis}{rgb}{.9,.0,.9}
\definecolor{christian}{rgb}{0,.6,.6}
\usepackage[normalem]{ulem}
\usepackage{cancel}

\newcommand{\ee}{\mathbbm{E}}

\begin{document}

\title{Quantum many-body systems out of equilibrium}

\author{J.\ Eisert, M.\ Friesdorf, and C.\ Gogolin}
\affiliation{Dahlem Center for Complex Quantum Systems, Freie Universit{\"a}t Berlin, 14195 Berlin, Germany}

\date{\today}

\begin{abstract}
Closed quantum many-body systems out of equilibrium pose several long-standing problems in physics. 
Recent years have seen a tremendous progress in approaching these questions, not least due to
experiments with cold atoms and trapped ions in instances of quantum simulations. 
This article provides an overview on the progress in understanding dynamical equilibration 
and thermalisation of closed quantum many-body systems out of equilibrium due to quenches, ramps and periodic driving. 
It also addresses topics such as the eigenstate thermalisation hypothesis, typicality, transport, many-body localisation, universality near phase transitions, and prospects for quantum simulations.
\end{abstract}

\maketitle

How do closed quantum many-body systems out of equilibrium eventually equilibrate? 
How and in precisely what way can such quantum systems with many degrees of freedom thermalise? 
It was established early on how to capture equilibrium states corresponding to some temperature in the framework 
of quantum statistical mechanics, but it seems much less clear how such states are eventually reached via the local dynamics following microscopic laws. 
This dichotomy gives rise to some tension between a microscopic description 
and one in terms of the familiar ensembles of statistical mechanics.
 It also leads to the question of how the actual dynamics of quantum phase transitions can be grasped. 
Asking questions of this type has a long tradition, but many problems have remained largely unresolved.
It was only relatively recently that they moved back into the focus of attention.
This development is driven by the availability of experiments that allow to probe such questions under controlled conditions, but 
also by new developments in theoretical physics. 
The present review aims at providing a first impression of important progress 
in this rapidly developing field of research, it emphasises promising perspectives, 
and collects a number of key references. Given its brevity, it can by no means be faithful to all developments or cover the whole literature of the field.

\section*{Local quantum many-body systems}
The physical systems considered here are \emph{local quantum many-body systems} with finite-range interactions, 
ubiquitous in condensed-matter physics. They can be described by some interaction graph, with physical constituents at the vertices. 
The system is governed by a Hamiltonian that can be written as
\begin{equation}
	H = \sum_j h_j.
\end{equation}
Most commonly considered are nearest-neighbour interactions,
where each $h_j$ acts non-trivially on finitely many sites.
Paradigmatic examples of this type are spin lattice systems, such as the Ising model,
as well as bosonic or fermionic lattice systems, prominently Fermi- and Bose-Hubbard models. In the latter 
case, the Hamiltonian takes the form
\begin{equation}
	H = J \sum_{\langle j,k\rangle }(b_j^\dagger b_k + b_k^\dagger b_j )+ U \sum_j n_j(n_j-1),
\end{equation}
with $n_j=b_j^\dagger b_j$, $b_j$ denoting the bosonic annihilation operator at site $j$,
the parameters $U,J$ being real and the sum $\langle j,k\rangle$ running over nearest neighbours only.
Common variants of this are models with other 
short range interactions, or long range interactions with a strength decaying polynomially in the distance.

\section*{Equilibration after quenches}
Investigating the fundamental connection between thermodynamics and the dynamics of closed quantum systems
has a long tradition\cite{vonNeumann29} but has also been a key topic in recent years.
A very clean setting for non-equilibrium dynamics is the one emerging from a sudden \emph{global quench}
\cite{Greiner,CalabreseCardy06,NumericalQuench,RigolFirst,AnalyticalQuench,CramerEisertScholl08,Rigol_etal08,Kehrein,Manmana2009,Polkovnikov_etal11,CalabreseEsslerFagotti11,Trotzky_etal12,CauxEssler,Barmettler}: 
Initially, the system is in a state $\rho_0$,
which could be the ground state of a local Hamiltonian.
Then one quickly alters the system's parameters globally and considers the many-body unitary time evolution 
under some local Hamiltonian $H$.
Of specific interest are expectation values of observables $A$
at later times 
\begin{equation}
  \langle A(t)\rangle = \tr(e^{-\i t H} \rho_0\, e^{\i t H} A) . 
\end{equation}
A main question is to what extent and for what times the situation can be described by a suitable equilibrium ensemble.
Notably, the dynamics is time reversal invariant and for finite systems recurrent, so a priori it seems far from
clear how and in what sense equilibrium can be reached dynamically.

A first important insight is that such systems generically indeed relax and equilibrate in the following sense: Even though the dynamics is 
entirely unitary, following transient non-equilibrium dynamics,
it is expectation values $\langle A(t)\rangle$ of \emph{many observables} that equilibrate. 
This is specifically true for \emph{local observables} which are supported only on a small number of sites. 
The apparent long time equilibrium state has to be equal to the time average
\begin{equation}
	\omega= \lim_{T\rightarrow\infty} \frac{1}{T}\int_0^T dt\, e^{-\i t H}\rho_0\, e^{\i t H}.
\end{equation}
This state is the \emph{maximum entropy state}, holding all constants of motion fixed\cite{Integrable}. 
In case of non-degenerate eigenvalues of the Hamiltonian, this ensemble is also called the
\emph{diagonal ensemble}\cite{GeneralizedGibbs}. 
This feature is reminiscent of a dynamical Jaynes' principle: 
A many-body system is pushed out of equilibrium and follows unitary dynamics. 
Yet, for most times, the systems appears as if it had equilibrated to a maximum entropy state.

The general expectation that many-body systems equilibrate can be made rigorous in a number of senses:
For some free models, specifically for the case of the integrable non-interacting limit of the Bose-Hubbard model,
equilibration is proven to be true in the strong sense of equilibration during intervals: 
Here, one can show for local observables $A$ that $|\langle A(t)\rangle - {\rm tr} (\omega A)|$ 
is arbitrarily small after a known initial relaxation time, and remains so for a time that grows linearly with the system size
\cite{AnalyticalQuench,CramerCLT}. 
One merely has to assume that the otherwise arbitrary initial state has suitable polynomially decaying correlations.
At the heart of such a rigorous argument are Lieb-Robinson bounds as well as non-commutative central limit theorems.

If no locality structure is available or is made use of, one can still show equilibration on average
for all Hamiltonians that have non-degenerate energy gaps: 
One can bound the time average $\ee_{{t \in [0,\infty)}} |\langle A(t)\rangle - {\rm tr} (\omega A)|$
by a quantity that is usually exponentially small in the system size\cite{Equilibration2,Linden_etal09}. This means that out of equilibrium quantum systems appear 
relaxed for overwhelmingly most times, even if by that statement alone no information on time scales can be deduced. More refined
bounds have been derived, also providing some information about the relevant time scales of 
equilibration\cite{ShortFarrelly12,ReimannKastner12}, but they are far off from those observed in 
numerical simulations\cite{NumericalQuench,CramerEisertScholl08,Barmettler,Polkovnikov_etal11} 
and experiments\cite{Trotzky_etal12,Langen2013,LightConeSchmiedmayer}. These findings are also related to further efforts
of describing the size of the time fluctuations after relaxation \cite{Lea,Zanardi}.
 
By no means are sudden global quenches the only relevant setting of quantum many-body systems out of equilibrium. 
Similarly important are scenarios of \emph{local quenches}\cite{LocalQuenches1,LocalQuenches2},
where not the entire system is uniformly modified, 
but rather the system is locally suddenly driven out of equilibrium 
by a change limited to a bounded number of lattice sites.
\emph{Ramps}\cite{Zurek_review,Polkovnikov_etal11,Emergence,Bakr,Haque_Crossover}, during which the Hamiltonian is changed according to some schedule, 
are of interest in studies of the dynamics of quantum phase transitions. 
The response of a system to a sudden alteration of its geometry can be studied by \emph{geometric quenches}\cite{GeometricCaux,GeometricQuench,RigolFirst}.
\emph{Time-dependent periodic driving} is discussed in a separate section. 
In this brief review, closed many-body systems are in the focus of attention.
It should be clear that 
specifically \emph{pump-probe} type settings\cite{PumpProbe} generate an out of equilibrium situation that can be used as a diagnostic tool in 
the condensed-matter context, giving rise to an interesting research topic in its own right.

\section*{Thermalisation}
The success of thermodynamics in describing large scale systems indicates that often 
the equilibrium values can be described by few parameters such as the global temperature and particle number.
We use the term \emph{thermalisation} to refer to the equilibration towards a state $\omega$
that is in a suitable sense close to being indistinguishable from a thermal equilibrium state proportional to $e^{-\beta H}$ for some inverse temperature $\beta>0$.
There are several mechanisms that can lead to thermalisation in this sense:
The first is based on the \emph{eigenstate thermalisation hypothesis} (ETH)\cite{Deutsch91,Srednicki94,tasaki98,Rigol_etal08,RigolETH}.
It conjectures that sufficiently complex quantum systems have eigenstates that --- for physically relevant observables such as local ones ---
are practically indistinguishable from thermal states with the same average energy.
The ETH and its breakdown in integrable systems
have recently been investigated extensively, mostly with numerical methods. Supporting evidence for it has been collected in a 
plethora of models\cite{Rigol_etal08,Rigol09,Gemmer_ETH,Polkovnikov_etal11,MoessnerETH}.

Without invoking the ETH one can still rigorously show thermalisation in weakly interacting systems under stronger conditions on the initial state\cite{Arnau}.
In addition, dynamical thermalisation of local expectation values to those in the global thermal state\cite{Mueller} has been shown under stronger conditions, 
most notably translation invariance and the existence of a unique Gibbs state. 
The notion of \emph{relative thermalisation}\cite{DelRio2014} 
focuses on the decoupling of a subsystem from a reference system.
So far, there is no compelling proof that 
globally quenched non-integrable systems dynamically thermalise.

These dynamical approaches are complemented by works based on \emph{typicality} arguments 
\cite{Goldstein06,Popescu06}.
Here, rather than following the dynamical evolution of a system, 
one tries to justify the applicability of ensembles by showing that 
most states drawn randomly according to some measure
have the same physical properties as some appropriate ensemble such as the canonical one.

If a quantum system is integrable\cite{Caux} in the sense that it has local conserved quantities, 
one should not expect the system to thermalise: These constants of motion prohibit full
thermalisation to the canonical ensemble. One can still expect the system to equilibrate to the maximum 
entropy state given these locally conserved quantities\cite{CauxEssler,Integrable}, a so called 
\emph{generalized Gibbs ensemble}\cite{GeneralizedGibbs,AnalyticalQuench}.
Effective thermalisation has been put into context with unitary quantum dynamics under conditions of classical chaos \cite{Altland}.

Several quantum many-body systems exhibit some instance of \emph{pre-thermalisation}:
This term refers to any apparent equilibration to a meta-stable state on a short time scale, before --- on a long time scale --- the system relaxes to a state
indistinguishable from a genuine thermal state. Examples are almost integrable systems\cite{PrethermalizationSilva,PrethermalisationEssler} and continuous systems of coupled Bose-Einstein condensates\cite{LightConeSchmiedmayer}.

\section*{``Light cones'' and entanglement dynamics} An important stepping stone for a better understanding of the non-equilibrium dynamics of local Hamiltonian models
is the insight that \emph{Lieb-Robinson bounds}\cite{HastingsKoma06,Nachtergaele_Locality,Quench2}
limit the speed of information propagation in such systems. There are several ways of stating such bounds: One is to say that for any two observables $A$ and $B$
\begin{equation}
	\| {[} A(t), B{]}\|\leq c\, \|A\|\, \|B\| \min \{|A|,|B|\}\, e^{-\mu\,(d(A,B)-v\,|t|)},
\end{equation}
where $d(A,B)$ is the distance between the support of the observables, $|A|$ and $|B|$ the size of their supports,
$v\geq 0$ takes the role of a group velocity,
and $c,\mu>0$ are constants. 
That is to say, information propagation outside a ``light-cone'' is exponentially suppressed. 
It immediately follows that correlation functions can significantly grow in time only inside the light cone \cite{Quench2}.
While these bounds are not necessarily tight, the picture that the local dynamics leads to excitations traveling through the system with velocity $v$
offers a lot of explanatory potential for non-equilibrium dynamics 
\cite{CalabreseCardy06,CalabreseQuenchEntanglement,AnalyticalQuench,CramerCLT,CramerEisertScholl08,Langen2013,Blatt_LR}.
Such light-cone dynamics has been experimentally observed in optical lattice
systems using novel tools that allow for single-site addressing\cite{1111.0776v1}
and in continuous systems of cold atoms\cite{LightConeSchmiedmayer}. 
  
The light-cone-like propagation of information has implications on \emph{entanglement dynamics}.
Many-body states are said to satisfy an \emph{area law} for the entanglement entropy, 
if for a subset of sites the latter scales only
like the boundary ``area'' of the subset rather than its ``volume''\cite{OneD,AreaReview}. 
It has been shown that if initially such a law holds, this will remain to be true for longer times\cite{Quench,Quench2}. 
Yet, at the same time, the pre-factor of this area law is expected to grow exponentially\cite{SchuchQuench}.
This is the ultimate reason why numerical \emph{tensor network methods}\cite{OrusReview,Folding}, such as the 
\emph{density-matrix renormalisation group method}\cite{Schollwock201196},
can simulate out of equilibrium dynamics efficiently for short times on a classical computer, 
while long times are not accessible. 
This is a feature shared by other powerful numerical methods as well, 
such as \emph{Monte-Carlo}\cite{TimeMonteCarlo} and \emph{dynamical mean field theory (DMFT)}
\cite{Eckstein_DMFT}.
This is in line with complexity-theoretic arguments that provide evidence that generic quantum long-time dynamics cannot be efficiently classically simulated \cite{JMathPhys510-1}.
If interactions are not strictly local, modified Lieb-Robinson bounds still hold 
\cite{HastingsKoma06},
giving rise to a rich phenomenology which has been experimentally explored in systems of trapped ions 
exhibiting long-range order\cite{Blatt_LR,1401.5088}.

\section*{Transport}
Understanding electronic transport is one of the main motivations that started the field of condensed matter physics and gave rise 
to a large body of work trying to accurately capture the conductivity properties of these systems.
Most relevant for the purpose of this review are the transport properties of simple paradigmatic models such as the Heisenberg spin chain
or Hubbard-type models.
In these models, transport is expected to lead to equilibration, as it allows the system to locally forget 
its precise initial configuration\cite{AnalyticalQuench,SecondReview}.
In spin chains, transport typically refers to a spreading of magnetised domains\cite{Langer_realtime}.
In optical lattices one commonly studies the transport of particles starting from an initially trapped situation
\cite{Schneider_fermionic_transport, Expansion}, or of quasi-particles\cite{1111.0776v1,Blatt_LR}. Two of the key aims are to distinguish between diffusive and ballistic transport 
and to understand the limitations of {\it linear response theory}\cite{Langer_realtime}.
While simple, integrable models can be solved exactly and thus provide an excellent benchmark, 
our understanding of transport in realistic setups heavily relies on numerical tools.
Tools explicitly storing the full wave function in the memory, such as exact diagonalisation,
are limited to comparably small systems.
Tensor network tools such t-DMRG, TEBD, and variants thereof \cite{OrusReview,Schollwock201196} 
allow to parametrize the state more efficiently and are accurate up to machine
precision as long as entanglement entropies are sufficiently small\cite{Schuch_MPS}.
Thus, for short and intermediate times these tools allow for an investigation of transport and equilibration 
in systems with several hundred sites\cite{Expansion,Langer_realtime,Trotzky_etal12,Barmettler}.

\section*{Absence of thermalisation and many-body localisation}
Crucial for transport properties of materials is the influence of irregularities such as defects.
The impact of random potentials on particle mobility has thus been the subject of intensive study over the last century, 
culminating in the famous result of Anderson, 
showing full localisation both of the eigenfunctions as well as of the dynamics
\cite{Anderson} for single particle \emph{disordered models}.
In the case of non-interacting fermions in one dimension or equivalently, via the Jordan-Wigner transformation, for free spin chains,
Anderson's result directly gives strong localisation of all fermionic modes.
This leads to the entanglement entropy remaining upper bounded from above \cite{Pollmann_unbounded} 
and a vanishing Lieb-Robinson velocity\cite{Sims_zeroLR}.
These systems thus fail to serve as their own heat bath and transport of any kind is completely suppressed.
For interacting many-body systems, the effects of disorder are still much less clear, despite great efforts 
\cite{Aleiner_Disorder}. 
Interestingly, both the support of time evolved local observables as well as the entanglement entropy are found to grow logarithmically in the 
interacting case and are thus unbounded\cite{Pollmann_unbounded}, in stark contrast to the non-interacting case.
Dynamical aspects of many-body localisation are expected to go along with localisation of the eigenstates.
In several recent studies localisation effects due to disorder have been associated with a lack of entanglement 
in many energy eigenstates\cite{HuseReview,Integrable,Bauer} 
or alternatively only in those below a so called \emph{mobility edge}\cite{Sims_zeroLR,Aleiner_Disorder}.
This could lead to a phase transition between a disordered insulator
with exactly zero conductivity and a conducting phase\cite{Aleiner_Disorder}.
The localisation of eigenstates leads to a clear violation of the ETH 
and connects to an absence of thermalisation effect for localising 
many-body systems 
and the existence of local constants of motion in these models\cite{Integrable,HuseReview}.
While these signatures of many-body localisation have been extensively explored, 
a comprehensive definition of many-body localisation is still lacking.

\section*{Dynamics of quantum phase transitions}
Exploring the dynamical signatures of phase transitions is an interesting problem in its own right. 
It connects to the important question of how slow an experimental ramp needs to be to avoid the creation of
defects when preparing a quantum phase and whether this is at all possible in the thermodynamic limit\cite{Polkovnikov08-1}.
A key theoretical model for these transitions, based on the adiabatic theorem,
has been provided by Kibble and Zurek\cite{Zurek_review} 
for the important case of moving from a gapped phase into criticality.
Based solely on the critical exponents of the model, this strikingly simple formalism predicts the number of introduced defects as a function of the velocity of the ramp.
It has originally been developed for thermal transitions, where it has also been experimentally tested
\cite{Kibble_Vortex,KZM_Ion,KZM_Ion2}. 
For quantum phase transitions such a scaling can be derived in the limit of infinitely slow ramps,  
based on universality arguments together with adiabatic perturbation theory\cite{Polkovnikov08-1}
and seems to be capable of accurately describing the dynamics of quantum phase transitions 
when crossing a single critical point in some models\cite{Zurek_review}.
For strongly correlated systems in realistic experimental settings, the dynamics of quantum phase transitions
is more elusive and there are indications that the complexity of the dynamics cannot
fully be captured by employing simple scaling arguments based only on the critical exponents of the model
\cite{Emergence, Chen_Slow}. The transition out of criticality has also been investigated, both numerically\cite{Kollath_Slow} and experimentally\cite{Bakr}.
Despite these first promising results, the dynamics of quantum phase transitions and how to capture them in
terms of simple scaling laws is still a largely open problem.

\section*{Quantum simulations}

When investigating the non-equilibrium behaviour of large-scale many-body systems,
analytical and numerical tools quickly reach their limits, especially for lattice models in more than one dimension.
A particularly exciting perspective arises from the idea that controlled quantum many-body systems constitute instances of 
\emph{analogue quantum simulators}\cite{CiracZollerSimulation,Trust,Trotzky_etal12,Emergence} overcoming these limitations.
These devices mimic natural interacting quantum many-body systems by reconstructing their Hamiltonian, but now under precisely 
controlled conditions. This is in contrast to \emph{digital quantum simulators}, anticipated devices that 
approximate quantum dynamics by suitable quantum circuits. Such analogue systems enable one to investigate 
some of the questions central to the research summarised in this review.

\emph{Ultra-cold atoms in optical lattices} provide a very promising platform for quantum simulators in this sense\cite{BlochSimulator}.
With present technology they can be used to investigate dynamics under Bose- and Fermi-Hubbard Hamiltonians
\cite{Greiner,Trotzky_etal12,Emergence,MonteCarloValidator,Esslinger2010,Koehl2014,Bakr,1111.0776v1,Schneider_fermionic_transport,Expansion,Chen_Slow} 
as well as for emulating spin chains\cite{Naegerl_Ising}. Due to the flexibility of optical lattice systems, e.g., by 
making use of Feshbach resonances, periodic driving, and superlattices, even sophisticated quantum simulations are conceivable, e.g., the simulation of 
\emph{lattice gauge theories}\cite{Gauge}.
Continuous optical setups allow for the investigation of out-of-equilibrium behaviour of quantum fields\cite{Hofferberth_etal07}
and have successfully been employed to investigate
(pre-)thermalisation\cite{LightConeSchmiedmayer} and out of equilibrium dynamics not leading to equilibration
on relevant time scales \cite{Kinoshita_etal06}.

Another promising architecture is provided by \emph{ion chains}\cite{BlattSimulator}, allowing for the precise control of the individual constituents\cite{Islam_etal11,KZM_Ion}.
Due to the involved Coulomb interaction, these systems are, among many other applications,
well suited to explore the validity of Lieb-Robinson bounds 
and their breakdown for long-range interactions\cite{Blatt_LR, 1401.5088}. 
\emph{Photonic architectures} as well as arrays of 
\emph{superconducting qubits} offer further promising platforms.
While the precise \emph{computational complexity} of analogue simulators has not been identified yet, and 
while it is unclear to what extent genuinely reliable analogue simulation is possible in the absence of fully-fledged quantum error correction, there is evidence that 
such analogue simulators have the potential of outperforming classical computers.
Quantum many-body dynamics can already be probed in experimental settings  under precisely controlled conditions for longer times
than can be kept track of using state-of-the-art tensor network methods on modern classical supercomputers\cite{Trotzky_etal12}.

\section*{Periodically driven systems}
Periodically driving a well controlled quantum system opens up new vistas for quantum simulations.
Consider a system with a Hamiltonian that satisfies $H(t)=H(t+\tau)$ for all times $t$ for some period $\tau>0$.
The eigenstates of such systems are usually constantly changing in time, but the long time dynamics can be inferred from their 
\emph{Floquet operator}\cite{PhysRevB.25.6622,Arimondo} $U(\tau)$, which is the time evolution operator over one period of driving. 
The Floquet operator gives rise to an effective Hamiltonian $H_\mathrm{eff}$ via
\begin{equation}
  U(\tau) = e^{-\i \tau H_\mathrm{eff}} ,
\end{equation}
and the time evolution of the driven system can be seen as a stroboscopic simulation\cite{PeriodicDriving3,PeriodicDriving2} of the time evolution under $H_\mathrm{eff}$.
The effective Hamiltonian can be tuned by choosing an appropriate driving schedule with great flexibility concerning the model parameters.
This allows the study of topological effects\cite{PeriodicDriving2} in driven systems and can be used to engineer gauge fields \cite{PeriodicDriving3}.
This has lead to an experimental realisation of the \emph{topological Haldane model} in a periodically modulated optical honeycomb lattice\cite{ExperimentFloquetDriving1} as well as
to experimental setups featuring \emph{Hofstadter bands}\cite{ExperimentFloquetDriving2} and \emph{spin-orbit coupling}
\cite{ExperimentFloquetDriving3} in optical lattice architectures \cite{Spielman}.
The synchronisation of the motion of periodically driven systems with the drive allows the description of the long time behavior by a \emph{periodic ensemble}.
For free systems it turns out to be given by a maximum entropy ensemble given the constants of motion\cite{Moessner} --- reminiscent of Jaynes' principle for systems in equilibrium and the generalised Gibbs ensemble emerging during equilibration after a quench.

\section*{Perspectives for open and closed systems}
The research field of quantum many-body systems out of equilibrium is rapidly developing and is 
taking new and unexpected directions at a remarkably high rate. This is partially driven by the impetus the field is subjected to through novel experimental 
developments. In optical lattice systems, \emph{single-site addressing}\cite{1111.0776v1,1303.5652,Bakr} 
opens up new possibilities for probing the out of equilibrium dynamics;
settings with periodic driving now allow for the study of topologically non-trivial situations\cite{PeriodicDriving2} and to develop a deeper understanding of particle physics and condensed matter systems\cite{PeriodicDriving3}.
\emph{Continuous systems} can be controlled with ever increasing precision \cite{Hofferberth_etal07,LightConeSchmiedmayer}, 
suggesting that such systems can serve as a platform for probing even effects of \emph{gravity}, 
such as the Unruh effect\cite{Unruh2,1402.6716}.
Hybrid architectures, such as cold atoms in optical lattices inside an optical cavity, constitute a particularly promising
avenue for experimental research\cite{ColdAtomsInCavities}.
Crucial for further progress will be a deeper understanding of the computational capability of analogue
quantum simulators and how their correct functioning can be certified in regimes that are classically not addressable\cite{Trotzky_etal12,MonteCarloValidator,Emergence,Trust}.

At the same time, new theoretical developments
keep the subject exciting, with \emph{quantum information theory} providing a fresh point of view and new methods.
\emph{Entanglement theory} can deliver important new insights specifically for the study of information 
propagation\cite{Quench,CalabreseQuenchEntanglement,AreaReview,OneD}.
\emph{Quantum thermodynamics}\cite{SmallestEngines} --- exploring the ultimate limits of thermodynamic 
processes in the quantum regime --- is now seeking contact with this field\cite{Gallego}, with interesting perspectives opening up. 
The \emph{AdS-cft-correspondence} also contributes a novel tool to study strongly correlated quantum systems out of equilibrium \cite{AdScft}.

In spite of the great progress, many questions are still wide open, even very basic ones. 
While it is understood under which conditions many-body systems
\emph{equilibrate}, it is far from clear on what \emph{time scales} they do so. 
The main reason for this is that the role of \emph{locality} for equilibration and thermalisation is, at least
from a rigorous perspective, far from being
understood. Steps in this direction have been taken only very recently
\cite{HastingsKoma06,Nachtergaele_Locality,Quench,Quench2,Mueller,Intensive}.
The conditions for \emph{thermalisation} are less clear and it is questionable whether (non-)integrability is able to capture this phenomenon in all of its facets\cite{Caux,Integrable}.
The problem of \emph{many-body localisation} has recently moved in the focus of attention \cite{HuseReview,Bauer},
but it seems fair to say that in the light of the various pictures of the phenomenon in the literature, 
a unifying framework is still missing.

A research area that deserves special attention is based on the idea that noise and dissipation 
are not necessarily detrimental to attempts to prepare and preserve interesting states of matter. 
In fact such preparations can be facilitated by suitably \emph{engineered local Markovian noise} \cite{DissipationZoller} that can be rigorously captured in
terms of a Liouvillian. 
In this mindset, an open system quantum simulator with trapped ions has been realised\cite{BlattOpen}, 
as well as one based of cold atoms controlled with localized dissipation
\cite{OttDissipation}.
In principle, even symmetry-protected topologically ordered states can be prepared by purely dissipative local Liouvillian dynamics\cite{DissipationTopologyZoller}.

It goes without saying that this perspectives article merely touches the surface of this exciting and rapidly developing field. Quantum many-body systems out of 
equilibrium allow us to probe long-standing questions in the foundations of statistical mechanics, relate to open questions in condensed-matter theory, and provide a promising platform for 
analogue quantum simulators outperforming classical computers.




We would like to thank many colleagues for numerous discussions over the years, and
E.\ Bergholtz,
E.\ T.\ Campbell,
A.\ del Campo,
J.-S.\ Caux, 
F.\ Essler,
T.\ Farrelly,
M.\ Rigol, 
and
R.\ Moessner
for comments on the manuscript,
and the EU (RAQUEL, SIQS), the ERC, 
the BMBF, and the Studienstiftung des Deutschen Volkes
for support.


\begin{thebibliography}{100}
\expandafter\ifx\csname url\endcsname\relax
  \def\url#1{\texttt{#1}}\fi
\expandafter\ifx\csname urlprefix\endcsname\relax\def\urlprefix{URL }\fi
\providecommand{\bibinfo}[2]{#2}
\providecommand{\eprint}[2][]{\url{#2}}

\bibitem{vonNeumann29}
\bibinfo{author}{v.~Neumann, J.}
\newblock \bibinfo{title}{{Beweis des Ergodensatzes und des H-Theorems in der
  neuen Mechanik}}.
\newblock \emph{\bibinfo{journal}{Z. Phys.}} \textbf{\bibinfo{volume}{57}},
  \bibinfo{pages}{30} (\bibinfo{year}{1929}).

\bibitem{Greiner}
\bibinfo{author}{Greiner, M.}, \bibinfo{author}{Mandel, O.},
  \bibinfo{author}{H{\"a}nsch, T.~W.} \& \bibinfo{author}{Bloch, I.}
\newblock \bibinfo{title}{{Collapse and revival of the matter wave field of a
  Bose-Einstein condensate}}.
\newblock \emph{\bibinfo{journal}{Nature}} \textbf{\bibinfo{volume}{419}},
  \bibinfo{pages}{51} (\bibinfo{year}{2002}).

\bibitem{CalabreseCardy06}
\bibinfo{author}{Calabrese, P.} \& \bibinfo{author}{Cardy, J.}
\newblock \bibinfo{title}{Time dependence of correlation functions following a
  quantum quench}.
\newblock \emph{\bibinfo{journal}{Phys. Rev. Lett.}}
  \textbf{\bibinfo{volume}{96}}, \bibinfo{pages}{136801}
  (\bibinfo{year}{2006}).

\bibitem{NumericalQuench}
\bibinfo{author}{Kollath, C.}, \bibinfo{author}{L{\"a}uchli, A.} \&
  \bibinfo{author}{Altman, E.}
\newblock \bibinfo{title}{{Quench dynamics and non equilibrium phase diagram of
  the Bose-Hubbard model}}.
\newblock \emph{\bibinfo{journal}{Phys. Rev. Lett.}}
  \textbf{\bibinfo{volume}{98}}, \bibinfo{pages}{180601}
  (\bibinfo{year}{2007}).

\bibitem{RigolFirst}
\bibinfo{author}{Rigol, M.}, \bibinfo{author}{Dunjko, V.},
  \bibinfo{author}{Yurovsky, V.} \& \bibinfo{author}{Olshanii, M.}
\newblock \bibinfo{title}{Relaxation in a completely integrable many-body
  quantum system: An ab initio study of the dynamics of the highly excited
  states of 1D lattice hard-core bosons}.
\newblock \emph{\bibinfo{journal}{Phys. Rev. Lett.}}
  \textbf{\bibinfo{volume}{98}}, \bibinfo{pages}{050405}
  (\bibinfo{year}{2007}).

\bibitem{AnalyticalQuench}
\bibinfo{author}{Cramer, M.}, \bibinfo{author}{Dawson, C.~M.},
  \bibinfo{author}{Eisert, J.} \& \bibinfo{author}{Osborne, T.~J.}
\newblock \bibinfo{title}{Quenching, relaxation, and a central limit theorem
  for quantum lattice systems}.
\newblock \emph{\bibinfo{journal}{Phys. Rev. Lett.}}
  \textbf{\bibinfo{volume}{100}}, \bibinfo{pages}{030602}
  (\bibinfo{year}{2008}).

\bibitem{CramerEisertScholl08}
\bibinfo{author}{Flesch, A.}, \bibinfo{author}{Cramer, M.},
  \bibinfo{author}{McCulloch, I.~P.}, \bibinfo{author}{Schollw{\"o}ck, U.} \&
  \bibinfo{author}{Eisert, J.}
\newblock \bibinfo{title}{Probing local relaxation of cold atoms in optical
  superlattices}.
\newblock \emph{\bibinfo{journal}{Phys. Rev. A}} \textbf{\bibinfo{volume}{78}},
  \bibinfo{pages}{033608} (\bibinfo{year}{2008}).

\bibitem{Kehrein}
	Moeckel, M. \& and Kehrein, S.
	Interaction quench in the Hubbard model.
	\emph{Phys. Rev. Lett.} {\bf 100}, 175702 (2008).
	
	
\bibitem{Rigol_etal08}
\bibinfo{author}{Rigol, M.}, \bibinfo{author}{Dunjko, V.} \&
  \bibinfo{author}{Olshanii, M.}
\newblock \bibinfo{title}{Thermalization and its mechanism for generic isolated
  quantum systems}.
\newblock \emph{\bibinfo{journal}{Nature}} \textbf{\bibinfo{volume}{452}},
  \bibinfo{pages}{854} (\bibinfo{year}{2008}).


\bibitem{Barthel}
Barthel, T., \& Schollw{\"o}ck, U.
\bibinfo{title}{Dephasing and the steady state in quantum many-particle systems.} 
\emph{Phys. Rev. Lett.} {\bf 100}, 100601 (2008).



\bibitem{Manmana2009}
\bibinfo{author}{Manmana, S.~R.}, \bibinfo{author}{Wessel, S.},
  \bibinfo{author}{Noack, R.~M.} \& \bibinfo{author}{Muramatsu, A.}
\newblock \bibinfo{title}{Time evolution of correlations in strongly
  interacting fermions after a quantum quench}.
\newblock \emph{\bibinfo{journal}{Phys. Rev. B}} \textbf{\bibinfo{volume}{79}},
  \bibinfo{pages}{155104} (\bibinfo{year}{2009}).

\bibitem{Polkovnikov_etal11}
\bibinfo{author}{Polkovnikov, A.}, \bibinfo{author}{Sengupta, K.},
  \bibinfo{author}{Silva, A.} \& \bibinfo{author}{Vengalattore, M.}
\newblock \bibinfo{title}{Non-equilibrium dynamics of closed interacting
  quantum systems}.
\newblock \emph{\bibinfo{journal}{Rev. Mod. Phys.}}
  \textbf{\bibinfo{volume}{83}}, \bibinfo{pages}{863} (\bibinfo{year}{2011}).

\bibitem{CalabreseEsslerFagotti11}
\bibinfo{author}{Calabrese, P.}, \bibinfo{author}{Essler, F. H.~L.} \&
  \bibinfo{author}{Fagotti, M.}
\newblock \bibinfo{title}{{Quantum quench in the transverse-field Ising
  chain}}.
\newblock \emph{\bibinfo{journal}{Phys. Rev. Lett.}}
  \textbf{\bibinfo{volume}{106}}, \bibinfo{pages}{227203}
  (\bibinfo{year}{2011}).

\bibitem{Trotzky_etal12}
\bibinfo{author}{Trotzky, S.} \emph{et~al.}
\newblock \bibinfo{title}{{Probing the relaxation towards equilibrium in an
  isolated strongly correlated one-dimensional Bose gas}}.
\newblock \emph{\bibinfo{journal}{Nature Phys.}} \textbf{\bibinfo{volume}{8}},
  \bibinfo{pages}{325} (\bibinfo{year}{2012}).

\bibitem{CauxEssler}
\bibinfo{author}{Caux, J.-S.} \& \bibinfo{author}{Essler, F. H.~L.}
\newblock \bibinfo{title}{Time evolution of local observables after quenching
  to an integrable model}.
\newblock \emph{\bibinfo{journal}{Phys. Rev. Lett.}}
  \textbf{\bibinfo{volume}{110}}, \bibinfo{pages}{257203}
  (\bibinfo{year}{2013}).

\bibitem{Barmettler}
\bibinfo{author}{Barmettler, P.}, \bibinfo{author}{Punk, M.},
  \bibinfo{author}{Gritsev, V.}, \bibinfo{author}{Demler, E.} \&
  \bibinfo{author}{Altman, E.}
\newblock \bibinfo{title}{Relaxation of antiferromagnetic order in spin-$\frac{1}{2}$
  chains following a quantum quench}.
\newblock \emph{\bibinfo{journal}{Phys. Rev. Lett.}}
  \textbf{\bibinfo{volume}{102}}, \bibinfo{pages}{130603}
  (\bibinfo{year}{2009}).

\bibitem{Integrable}
\bibinfo{author}{Gogolin, C.}, \bibinfo{author}{Mueller, M.~P.} \&
  \bibinfo{author}{Eisert, J.}
\newblock \bibinfo{title}{Absence of thermalization in non-integrable systems}.
\newblock \emph{\bibinfo{journal}{Phys. Rev. Lett.}}
  \textbf{\bibinfo{volume}{106}}, \bibinfo{pages}{040401}
  (\bibinfo{year}{2011}).

\bibitem{GeneralizedGibbs}
\bibinfo{author}{Cassidy, A.~C.}, \bibinfo{author}{Clark, C.~W.} \&
  \bibinfo{author}{Rigol, M.}
\newblock \bibinfo{title}{Generalized thermalization in an integrable lattice
  system}.
\newblock \emph{\bibinfo{journal}{Phys. Rev. Lett.}}
  \textbf{\bibinfo{volume}{106}}, \bibinfo{pages}{140405}
  (\bibinfo{year}{2011}).

\bibitem{CramerCLT}
\bibinfo{author}{Cramer, M.} \& \bibinfo{author}{Eisert, J.}
\newblock \bibinfo{title}{A quantum central limit theorem for non-equilibrium
  systems: Exact local relaxation of correlated states}.
\newblock \emph{\bibinfo{journal}{New J. Phys.}} \textbf{\bibinfo{volume}{12}},
  \bibinfo{pages}{055020} (\bibinfo{year}{2010}).

\bibitem{Equilibration2}
\bibinfo{author}{Reimann, P.}
\newblock \bibinfo{title}{Foundation of statistical mechanics under
  experimentally realistic conditions}.
\newblock \emph{\bibinfo{journal}{Phys. Rev. Lett.}}
  \textbf{\bibinfo{volume}{101}}, \bibinfo{pages}{190403}
  (\bibinfo{year}{2008}).

\bibitem{Linden_etal09}
\bibinfo{author}{Linden, N.}, \bibinfo{author}{Popescu, S.},
  \bibinfo{author}{Short, A.~J.} \& \bibinfo{author}{Winter, A.}
\newblock \bibinfo{title}{Quantum mechanical evolution towards thermal
  equilibrium}.
\newblock \emph{\bibinfo{journal}{Phys. Rev. E}} \textbf{\bibinfo{volume}{79}},
  \bibinfo{pages}{061103} (\bibinfo{year}{2009}).

\bibitem{ReimannKastner12}
\bibinfo{author}{Reimann, P.} \& \bibinfo{author}{Kastner, M.}
\newblock \bibinfo{title}{Equilibration of isolated macroscopic quantum
  systems}.
\newblock \emph{\bibinfo{journal}{New J. Phys.}} \textbf{\bibinfo{volume}{14}},
  \bibinfo{pages}{043020} (\bibinfo{year}{2012}).

\bibitem{ShortFarrelly12}
\bibinfo{author}{Short, A.~J.} \& \bibinfo{author}{Farrelly, T.~C.}
\newblock \bibinfo{title}{Quantum equilibration in finite time}.
\newblock \emph{\bibinfo{journal}{New J. Phys.}} \textbf{\bibinfo{volume}{14}},
  \bibinfo{pages}{013063} (\bibinfo{year}{2012}).

\bibitem{Langen2013}
\bibinfo{author}{Langen, T.}, \bibinfo{author}{Geiger, R.},
  \bibinfo{author}{Kuhnert, M.}, \bibinfo{author}{Rauer, B.} \&
  \bibinfo{author}{Schmiedmayer, J.}
\newblock \bibinfo{title}{{Local emergence of thermal correlations in an
  isolated quantum many-body system}}.
\newblock \emph{\bibinfo{journal}{Nature Phys.}} \textbf{\bibinfo{volume}{9}},
  \bibinfo{pages}{640} (\bibinfo{year}{2013}).

\bibitem{LightConeSchmiedmayer}
\bibinfo{author}{Geiger, R.}, \bibinfo{author}{Langen, T.},
  \bibinfo{author}{Mazets, I.} \& \bibinfo{author}{Schmiedmayer, J.}
\newblock \bibinfo{title}{Local relaxation and light-cone-like propagation of
  correlations in a trapped one-dimensional Bose gas}.
\newblock \emph{\bibinfo{journal}{New. J. Phys.}}
  \textbf{\bibinfo{volume}{16}}, \bibinfo{pages}{053034}
  (\bibinfo{year}{2014}).

\bibitem{Lea}
\bibinfo{author}{Torres-Herrera, E.~J.}, \bibinfo{author}{Kollmar, D.} \&
  \bibinfo{author}{Santos, L.~F.}
\newblock \bibinfo{title}{Relaxation and thermalization of isolated
  many-body quantum systems}.
\newblock \bibinfo{note}{ArXiv:1403.6481}.

\bibitem{Zanardi}
\bibinfo{author}{Venuti, L.~C.} \& \bibinfo{author}{Zanardi, P.}
\newblock \bibinfo{title}{Universal time-fluctuations in near-critical
  out-of-equilibrium quantum dynamics}.
\newblock \emph{\bibinfo{journal}{Phys. Rev. E}} \textbf{\bibinfo{volume}{89}},
  \bibinfo{pages}{022101} (\bibinfo{year}{2014}).

\bibitem{LocalQuenches1}
\bibinfo{author}{Torres-Herrera, E.~J.} \& \bibinfo{author}{Santos, L.~F.}
\newblock \bibinfo{title}{Local quenches with global effects in interacting
  quantum systems}.
\newblock \emph{\bibinfo{journal}{Phys. Rev. E}} \textbf{\bibinfo{volume}{89}},
  \bibinfo{pages}{062110} (\bibinfo{year}{2014}).

\bibitem{LocalQuenches2}
\bibinfo{author}{Stephan, J.-M.} \& \bibinfo{author}{Dubail, J.}
\newblock \bibinfo{title}{{Local quantum quenches in critical one-dimensional
  systems: Entanglement, the Loschmidt echo, and light-cone effects}}.
\newblock \emph{\bibinfo{journal}{J. Stat. Mech.}} \bibinfo{pages}{P08019}
  (\bibinfo{year}{2011}).

\bibitem{Zurek_review}
\bibinfo{author}{del Campo, A.} \& \bibinfo{author}{Zurek, W.~H.}
\newblock \bibinfo{title}{Universality of phase transition dynamics:
  Topological defects from symmetry breaking}.
\newblock \emph{\bibinfo{journal}{Int. J. Mod. Phys. A}}
  \textbf{\bibinfo{volume}{29}}, \bibinfo{pages}{1430018}
  (\bibinfo{year}{2014}).

\bibitem{Emergence}
\bibinfo{author}{Braun, S.} \emph{et~al.}
\newblock \bibinfo{title}{Emergence of coherence and the dynamics of quantum
  phase transitions}.
\newblock \bibinfo{note}{ArXiv:1403.7199}.

\bibitem{Bakr}
\bibinfo{author}{Bakr, W.~S.} \emph{et~al.}
\newblock \bibinfo{title}{{Probing the superfluid-to-Mott insulator transition
  at the single-atom level}}.
\newblock \emph{\bibinfo{journal}{Science}} \textbf{\bibinfo{volume}{329}},
  \bibinfo{pages}{547} (\bibinfo{year}{2010}).

\bibitem{Haque_Crossover}
\bibinfo{author}{D\'ora, B.}, \bibinfo{author}{Haque, M.} \&
  \bibinfo{author}{Zar\'and, G.}
\newblock \bibinfo{title}{{Crossover from adiabatic to sudden interaction
  quench in a Luttinger liquid}}.
\newblock \emph{\bibinfo{journal}{Phys. Rev. Lett.}}
  \textbf{\bibinfo{volume}{106}}, \bibinfo{pages}{156406}
  (\bibinfo{year}{2011}).

\bibitem{GeometricCaux}
\bibinfo{author}{Mossel, J.}, \bibinfo{author}{Palacios, G.} \&
  \bibinfo{author}{Caux, J.-S.}
\newblock \bibinfo{title}{Geometric quenches in quantum integrable
  systems}.
\newblock \emph{\bibinfo{journal}{J. Stat. Mech.}} \bibinfo{pages}{L09001}
  (\bibinfo{year}{2010}).

\bibitem{GeometricQuench}
\bibinfo{author}{Alba, V.} \& \bibinfo{author}{Heidrich-Meisner, F.}
\newblock \bibinfo{title}{Entanglement spreading after a geometric quench in
  quantum spin chains}.
\newblock \bibinfo{note}{ArXiv:1402.2299}.

\bibitem{PumpProbe}
\bibinfo{editor}{Bovensiepen, U.}, \bibinfo{editor}{Petek, H.} \&
  \bibinfo{editor}{Wolf, M.} (eds.) \emph{\bibinfo{title}{Dynamics at solid
  state surfaces and interfaces}} (\bibinfo{publisher}{Wiley-VCH},
  \bibinfo{address}{Weinheim, Germany}, \bibinfo{year}{2010}).

\bibitem{Deutsch91}
\bibinfo{author}{Deutsch, J.~M.}
\newblock \bibinfo{title}{Quantum statistical mechanics in a closed system}.
\newblock \emph{\bibinfo{journal}{Phys. Rev. A}} \textbf{\bibinfo{volume}{43}},
  \bibinfo{pages}{2046} (\bibinfo{year}{1991}).

\bibitem{Srednicki94}
\bibinfo{author}{Srednicki, M.}
\newblock \bibinfo{title}{Chaos and quantum thermalization}.
\newblock \emph{\bibinfo{journal}{Phys. Rev. E}} \textbf{\bibinfo{volume}{50}},
  \bibinfo{pages}{888} (\bibinfo{year}{1994}).

\bibitem{tasaki98}
\bibinfo{author}{Tasaki, H.}
\newblock \bibinfo{title}{From quantum dynamics to the canonical distribution:
  General picture and a rigorous example}.
\newblock \emph{\bibinfo{journal}{Phys. Rev. Lett.}}
  \textbf{\bibinfo{volume}{80}}, \bibinfo{pages}{1373} (\bibinfo{year}{1998}).

\bibitem{RigolETH}
\bibinfo{author}{Rigol, M.} \& \bibinfo{author}{Srednicki, M.}
\newblock \bibinfo{title}{Alternatives to eigenstate thermalization}.
\newblock \emph{\bibinfo{journal}{Phys. Rev. Lett.}}
  \textbf{\bibinfo{volume}{108}}, \bibinfo{pages}{110601}
  (\bibinfo{year}{2012}).

\bibitem{Rigol09}
\bibinfo{author}{Rigol, M.}
\newblock \bibinfo{title}{Breakdown of thermalization in finite one-dimensional
  systems}.
\newblock \emph{\bibinfo{journal}{Phys. Rev. Lett.}}
  \textbf{\bibinfo{volume}{103}}, \bibinfo{pages}{100403}
  (\bibinfo{year}{2009}).

\bibitem{Gemmer_ETH}
\bibinfo{author}{Steinigeweg, R.}, \bibinfo{author}{Khodja, A.},
  \bibinfo{author}{Niemeyer, H.}, \bibinfo{author}{Gogolin, C.} \&
  \bibinfo{author}{Gemmer, J.}
\newblock \bibinfo{title}{Pushing the limits of the eigenstate thermalization
  hypothesis towards mesoscopic quantum systems}.
\newblock \emph{\bibinfo{journal}{Phys. Rev. Lett.}}
  \textbf{\bibinfo{volume}{112}}, \bibinfo{pages}{130403}
  (\bibinfo{year}{2014}).

\bibitem{MoessnerETH}
\bibinfo{author}{Beugeling, W.}, \bibinfo{author}{Moessner, R.} \&
  \bibinfo{author}{Haque, M.}
\newblock \bibinfo{title}{Finite-size scaling of eigenstate thermalization}.
\newblock \emph{\bibinfo{journal}{Phys. Rev. E}} \textbf{\bibinfo{volume}{89}},
  \bibinfo{pages}{042112} (\bibinfo{year}{2014}).

\bibitem{Arnau}
\bibinfo{author}{Riera, A.}, \bibinfo{author}{Gogolin, C.} \&
  \bibinfo{author}{Eisert, J.}
\newblock \bibinfo{title}{Thermalization in nature and on a quantum computer}.
\newblock \emph{\bibinfo{journal}{Phys. Rev. Lett.}}
  \textbf{\bibinfo{volume}{108}}, \bibinfo{pages}{080402}
  (\bibinfo{year}{2012}).

\bibitem{Mueller}
\bibinfo{author}{Mueller, M.~P.}, \bibinfo{author}{Adlam, E.},
  \bibinfo{author}{Masanes, L.} \& \bibinfo{author}{Wiebe, N.}
\newblock \bibinfo{title}{Thermalization and canonical typicality in
  translation-invariant quantum lattice systems}.
\newblock \bibinfo{note}{ArXiv:1312.7420}.

\bibitem{DelRio2014}
\bibinfo{author}{del Rio, L.}, \bibinfo{author}{Hutter, A.},
  \bibinfo{author}{Renner, R.} \& \bibinfo{author}{Wehner, S.}
\newblock \bibinfo{title}{Relative thermalization} (\bibinfo{year}{2014}).
\newblock \bibinfo{note}{ArXiv:1401.7997}.

\bibitem{Goldstein06}
\bibinfo{author}{Goldstein, S.}
\newblock \bibinfo{title}{Canonical typicality}.
\newblock \emph{\bibinfo{journal}{Phys. Rev. Lett.}}
  \textbf{\bibinfo{volume}{96}}, \bibinfo{pages}{050403}
  (\bibinfo{year}{2006}).

\bibitem{Popescu06}
\bibinfo{author}{Popescu, S.}, \bibinfo{author}{Short, A.~J.} \&
  \bibinfo{author}{Winter, A.}
\newblock \bibinfo{title}{The foundations of statistical mechanics from
  entanglement: Individual states vs. averages}.
\newblock \emph{\bibinfo{journal}{Nature Phys.}} \textbf{\bibinfo{volume}{2}},
  \bibinfo{pages}{754} (\bibinfo{year}{2006}).

\bibitem{Caux}
\bibinfo{author}{Caux, J.-S.} \& \bibinfo{author}{Mossel, J.}
\newblock \bibinfo{title}{Remarks on the notion of quantum integrability}.
\newblock \emph{\bibinfo{journal}{J. Stat. Mech.}} \bibinfo{pages}{P02023}
  (\bibinfo{year}{2011}).

\bibitem{Altland}
\bibinfo{author}{Altland, A.} \& \bibinfo{author}{Haake, F.}
\newblock \bibinfo{title}{Quantum chaos and effective thermalization}.
\newblock \emph{\bibinfo{journal}{Phys. Rev. Lett.}}
  \textbf{\bibinfo{volume}{108}}, \bibinfo{pages}{073601}
  (\bibinfo{year}{2012}).

\bibitem{PrethermalizationSilva}
\bibinfo{author}{Marcuzzi, M.}, \bibinfo{author}{Marino, J.},
  \bibinfo{author}{Gambassi, A.} \& \bibinfo{author}{Silva, A.}
\newblock \bibinfo{title}{Prethermalization in a non-integrable quantum spin
  chain after a quench}.
\newblock \emph{\bibinfo{journal}{Phys. Rev. Lett.}}
  \textbf{\bibinfo{volume}{111}}, \bibinfo{pages}{197203}
  (\bibinfo{year}{2013}).

\bibitem{PrethermalisationEssler}
\bibinfo{author}{Essler, F.~H.~L.}, 
\bibinfo{author}{Kehrein, S.},
\bibinfo{author}{Manmana, S.~R.},
\bibinfo{author}{Robinson, N.~J.},
\newblock \bibinfo{title}{Quench dynamics in a model with tuneable integrability breaking}.
\newblock \emph{\bibinfo{journal}{Phys. Rev. B}}
  \textbf{\bibinfo{volume}{89}}, \bibinfo{pages}{165104}
  (\bibinfo{year}{2014}).

\bibitem{HastingsKoma06}
\bibinfo{author}{Hastings, M.~B.} \& \bibinfo{author}{Koma, T.}
\newblock \bibinfo{title}{Spectral gap and exponential decay of correlations}.
\newblock \emph{\bibinfo{journal}{Commun. Math. Phys.}}
  \textbf{\bibinfo{volume}{265}}, \bibinfo{pages}{781} (\bibinfo{year}{2006}).

\bibitem{Nachtergaele_Locality}
\bibinfo{editor}{Sidoravicius, V.} (ed.).
\newblock \emph{\bibinfo{title}{Locality estimates for quantum spin systems}}
  (\bibinfo{publisher}{Springer Verlag}, \bibinfo{year}{2009}).

\bibitem{Quench2}
\bibinfo{author}{Bravyi, S.}, \bibinfo{author}{Hastings, M.~B.} \&
  \bibinfo{author}{Verstraete, F.}
\newblock \bibinfo{title}{{Lieb-Robinson bounds and the generation of
  correlations and topological quantum order}}.
\newblock \emph{\bibinfo{journal}{Phys. Rev. Lett.}}
  \textbf{\bibinfo{volume}{97}}, \bibinfo{pages}{050401}
  (\bibinfo{year}{2006}).

\bibitem{CalabreseQuenchEntanglement}
\bibinfo{author}{Calabrese, P.} \& \bibinfo{author}{Cardy, J.}
\newblock \bibinfo{title}{Evolution of entanglement entropy in one-dimensional
  systems}.
\newblock \emph{\bibinfo{journal}{J. Stat. Mech.}} \bibinfo{pages}{P04010}
  (\bibinfo{year}{2005}).

\bibitem{Blatt_LR}
\bibinfo{author}{Jurcevic, P.} \emph{et~al.}
\newblock \bibinfo{title}{Quasiparticle engineering and entanglement
  propagation in a quantum many-body system}.
\newblock \emph{\bibinfo{journal}{Nature}} \textbf{\bibinfo{volume}{511}},
  \bibinfo{pages}{202} (\bibinfo{year}{2014}).

\bibitem{1111.0776v1}
\bibinfo{author}{Cheneau, M.} \emph{et~al.}
\newblock \bibinfo{title}{{Light-cone-like spreading of correlations in a
  quantum many-body system}}.
\newblock \emph{\bibinfo{journal}{Nature}} \textbf{\bibinfo{volume}{481}},
  \bibinfo{pages}{484} (\bibinfo{year}{2012}).

\bibitem{OneD}
\bibinfo{author}{Hastings, M.~B.}
\newblock \bibinfo{title}{An area law for one-dimensional quantum systems}.
\newblock \emph{\bibinfo{journal}{J. Stat. Mech.}} \bibinfo{pages}{P08024}
  (\bibinfo{year}{2007}).

\bibitem{AreaReview}
\bibinfo{author}{Eisert, J.}, \bibinfo{author}{Cramer, M.} \&
  \bibinfo{author}{Plenio, M.~B.}
\newblock \bibinfo{title}{Area laws for the entanglement entropy}.
\newblock \emph{\bibinfo{journal}{Rev. Mod. Phys.}}
  \textbf{\bibinfo{volume}{82}}, \bibinfo{pages}{277} (\bibinfo{year}{2010}).

\bibitem{Quench}
\bibinfo{author}{Eisert, J.} \& \bibinfo{author}{Osborne, T.~J.}
\newblock \bibinfo{title}{General entanglement scaling laws from time
  evolution}.
\newblock \emph{\bibinfo{journal}{Phys. Rev. Lett}}
  \textbf{\bibinfo{volume}{97}}, \bibinfo{pages}{150404}
  (\bibinfo{year}{2006}).

\bibitem{SchuchQuench}
\bibinfo{author}{Schuch, N.}, \bibinfo{author}{Wolf, M.~M.},
  \bibinfo{author}{Vollbrecht, K. G.~H.} \& \bibinfo{author}{Cirac, J.~I.}
\newblock \bibinfo{title}{On entropy growth and the hardness of simulating time
  evolution}.
\newblock \emph{\bibinfo{journal}{New J. Phys.}} \textbf{\bibinfo{volume}{10}},
  \bibinfo{pages}{033032} (\bibinfo{year}{2008}).

\bibitem{OrusReview}
\bibinfo{author}{Orus, R.}
\newblock \bibinfo{title}{A practical introduction to tensor networks: Matrix
  product states and projected entangled pair states}.
\newblock \emph{\bibinfo{journal}{Ann. Phys.}} \textbf{\bibinfo{volume}{349}},
  \bibinfo{pages}{117} (\bibinfo{year}{2014}).

\bibitem{Folding}
\bibinfo{author}{Banuls, M.~C.}, \bibinfo{author}{Hastings, M.~B.},
  \bibinfo{author}{Verstraete, F.} \& \bibinfo{author}{Cirac, J.~I.}
\newblock \bibinfo{title}{Matrix product states for dynamical simulation of
  infinite chains}.
\newblock \emph{\bibinfo{journal}{Phys. Rev. Lett.}}
  \textbf{\bibinfo{volume}{102}}, \bibinfo{pages}{240603}
  (\bibinfo{year}{2009}).

\bibitem{Schollwock201196}
\bibinfo{author}{Schollw{\"o}ck, U.}
\newblock \bibinfo{title}{The density-matrix renormalization group in the age
  of matrix product states}.
\newblock \emph{\bibinfo{journal}{Ann. Phys.}} \textbf{\bibinfo{volume}{326}},
  \bibinfo{pages}{96} (\bibinfo{year}{2011}).

\bibitem{TimeMonteCarlo}
\bibinfo{author}{Troyer, M.}, \bibinfo{author}{Alet, F.},
  \bibinfo{author}{Trebst, S.} \& \bibinfo{author}{Wessel, S.}
\newblock \bibinfo{title}{{Non-local updates for quantum Monte Carlo
  simulations}}.
\newblock \emph{\bibinfo{journal}{AIP Conf. Proc.}}
  \textbf{\bibinfo{volume}{690}}, \bibinfo{pages}{156} (\bibinfo{year}{2003}).

\bibitem{Eckstein_DMFT}
\bibinfo{author}{Aoki, H.} \emph{et~al.}
\newblock \bibinfo{title}{Non-equilibrium dynamical mean-field theory and its
  applications}.
\newblock \emph{\bibinfo{journal}{Rev. Mod. Phys.}}
  \textbf{\bibinfo{volume}{86}}, \bibinfo{pages}{779} (\bibinfo{year}{2014}).

\bibitem{JMathPhys510-1}
\bibinfo{author}{Nagaj, D.}
\newblock \bibinfo{title}{{Fast universal quantum computation with
  railroad-switch local Hamiltonians}}.
\newblock \emph{\bibinfo{journal}{J. Math. Phys.}}
  \textbf{\bibinfo{volume}{51}}, \bibinfo{pages}{62201} (\bibinfo{year}{2010}).

\bibitem{1401.5088}
\bibinfo{author}{Richerme, P.} \emph{et~al.}
\newblock \bibinfo{title}{Non-local propagation of correlations in long-range
  interacting quantum systems} (\bibinfo{year}{2014}).
\newblock \bibinfo{note}{ArXiv:1401.5088}.

\bibitem{SecondReview}
\bibinfo{author}{Gogolin, C.} \& \bibinfo{author}{Eisert, J.}
\newblock \bibinfo{title}{Pure state statistical mechanics}.
\newblock \emph{\bibinfo{journal}{In preparation}}  (\bibinfo{year}{2014}).

\bibitem{Langer_realtime}
\bibinfo{author}{Langer, S.}, \bibinfo{author}{Heidrich-Meisner, F.},
  \bibinfo{author}{Gemmer, J.}, \bibinfo{author}{McCulloch, I.~P.} \&
  \bibinfo{author}{Schollw\"ock, U.}
\newblock \bibinfo{title}{Real-time study of diffusive and ballistic transport
  in spin-$\frac{1}{2}$ chains using the adaptive time-dependent density matrix
  renormalization group method}.
\newblock \emph{\bibinfo{journal}{Phys. Rev. B}} \textbf{\bibinfo{volume}{79}},
  \bibinfo{pages}{214409} (\bibinfo{year}{2009}).

\bibitem{Schneider_fermionic_transport}
\bibinfo{author}{Schneider, U.} \emph{et~al.}
\newblock \bibinfo{title}{{Fermionic transport and out-of-equilibrium dynamics
  in a homogeneous Hubbard model with ultra-cold atoms}}.
\newblock \emph{\bibinfo{journal}{Nat. Phys.}} \textbf{\bibinfo{volume}{8}},
  \bibinfo{pages}{213} (\bibinfo{year}{2012}).

\bibitem{Expansion}
\bibinfo{author}{Ronzheimer, J.~P.} \emph{et~al.}
\newblock \bibinfo{title}{Expansion dynamics of interacting bosons in
  homogeneous lattices in one and two dimensions}.
\newblock \emph{\bibinfo{journal}{Phys. Rev. Lett.}}
  \textbf{\bibinfo{volume}{110}}, \bibinfo{pages}{205301}
  (\bibinfo{year}{2013}).

\bibitem{Schuch_MPS}
\bibinfo{author}{Schuch, N.}, \bibinfo{author}{Wolf, M.~M.},
  \bibinfo{author}{Verstraete, F.} \& \bibinfo{author}{Cirac, J.~I.}
\newblock \bibinfo{title}{Entropy scaling and simulability by matrix product
  states}.
\newblock \emph{\bibinfo{journal}{Phys. Rev. Lett.}}
  \textbf{\bibinfo{volume}{100}}, \bibinfo{pages}{030504}
  (\bibinfo{year}{2008}).

\bibitem{Anderson}
\bibinfo{author}{Lagendijk, A.}, \bibinfo{author}{van Tiggelen, B.} \&
  \bibinfo{author}{Wiersma, D.~S.}
\newblock \bibinfo{title}{{Fifty years of Anderson localization}}.
\newblock \emph{\bibinfo{journal}{Physics Today}}
  \textbf{\bibinfo{volume}{62}}, \bibinfo{pages}{24} (\bibinfo{year}{2009}).

\bibitem{Pollmann_unbounded}
\bibinfo{author}{Bardarson, J.~H.}, \bibinfo{author}{Pollmann, F.} \&
  \bibinfo{author}{Moore, J.~E.}
\newblock \bibinfo{title}{Unbounded growth of entanglement in models of
  many-body localization}.
\newblock \emph{\bibinfo{journal}{Phys. Rev. Lett.}}
  \textbf{\bibinfo{volume}{109}}, \bibinfo{pages}{017202}
  (\bibinfo{year}{2012}).

\bibitem{Sims_zeroLR}
\bibinfo{author}{Hamza, E.}, \bibinfo{author}{Sims, R.} \&
  \bibinfo{author}{Stolz, G.}
\newblock \bibinfo{title}{Dynamical localization in disordered quantum spin
  systems}.
\newblock \emph{\bibinfo{journal}{Commun. Math. Phys.}}
  \textbf{\bibinfo{volume}{315}}, \bibinfo{pages}{215} (\bibinfo{year}{2012}).

\bibitem{Aleiner_Disorder}
\bibinfo{author}{Aleiner, I.~L.}, \bibinfo{author}{Altshuler, B.~L.} \&
  \bibinfo{author}{Shlyapnikov, G.~V.}
\newblock \bibinfo{title}{A finite-temperature phase transition for disordered
  weakly interacting bosons in one dimension}.
\newblock \emph{\bibinfo{journal}{Ann. Phys.}} \textbf{\bibinfo{volume}{321}},
  \bibinfo{pages}{1126} (\bibinfo{year}{2006}).

\bibitem{HuseReview}
\bibinfo{author}{Nandkishore, R.} \& \bibinfo{author}{Huse, D.~A.}
\newblock \bibinfo{title}{Many body localization and thermalization in quantum
  statistical mechanics}.
\newblock \bibinfo{note}{ArXiv:1404.0686}.

\bibitem{Bauer}
\bibinfo{author}{Bauer, B.} \& \bibinfo{author}{Nayak, C.}
\newblock \bibinfo{title}{Area laws in a many-body localized state and its
  implications for topological order}.
\newblock \emph{\bibinfo{journal}{J. Stat. Mech.}} \bibinfo{pages}{P09005}
  (\bibinfo{year}{2013}).

\bibitem{Polkovnikov08-1}
\bibinfo{author}{Polkovnikov, A.} \& \bibinfo{author}{Gritsev, V.}
\newblock \bibinfo{title}{{Breakdown of the adiabatic limit in low-dimensional
  gapless systems}}.
\newblock \emph{\bibinfo{journal}{Nature Phys.}} \textbf{\bibinfo{volume}{4}},
  \bibinfo{pages}{477} (\bibinfo{year}{2008}).

\bibitem{Kibble_Vortex}
\bibinfo{author}{Ruutu, V. M.~H.} \emph{et~al.}
\newblock \bibinfo{title}{{Vortex formation in neutron-irradiated superfluid
  3He as an analogue of cosmological defect formation}}.
\newblock \emph{\bibinfo{journal}{Nature}} \textbf{\bibinfo{volume}{382}},
  \bibinfo{pages}{334} (\bibinfo{year}{1996}).

\bibitem{KZM_Ion}
\bibinfo{author}{Ulm, S.} \emph{et~al.}
\newblock \bibinfo{title}{{Observation of the Kibble-Zurek scaling law for
  defect formation in ion crystals}}.
\newblock \emph{\bibinfo{journal}{Nat. Comm.}} \textbf{\bibinfo{volume}{4}},
  \bibinfo{pages}{2290} (\bibinfo{year}{2013}).

\bibitem{KZM_Ion2}
\bibinfo{author}{Pyka, K.} \emph{et~al.}
\newblock \bibinfo{title}{Topological defect formation and spontaneous symmetry
  breaking in ion coulomb crystals}.
\newblock \emph{\bibinfo{journal}{Nat. Comm.}} \textbf{\bibinfo{volume}{4}}
  (\bibinfo{year}{2013}).

\bibitem{Chen_Slow}
\bibinfo{author}{Chen, D.}, \bibinfo{author}{White, M.},
  \bibinfo{author}{Borries, C.} \& \bibinfo{author}{DeMarco, B.}
\newblock \bibinfo{title}{{Quantum quench of an atomic Mott insulator}}.
\newblock \emph{\bibinfo{journal}{Phys. Rev. Lett.}}
  \textbf{\bibinfo{volume}{106}}, \bibinfo{pages}{235304}
  (\bibinfo{year}{2011}).

\bibitem{Kollath_Slow}
\bibinfo{author}{Bernier, J.-S.}, \bibinfo{author}{Poletti, D.},
  \bibinfo{author}{Barmettler, P.}, \bibinfo{author}{Roux, G.} \&
  \bibinfo{author}{Kollath, C.}
\newblock \bibinfo{title}{{Slow quench dynamics of Mott-insulating regions in a
  trapped Bose gas}}.
\newblock \emph{\bibinfo{journal}{Phys. Rev. A}} \textbf{\bibinfo{volume}{85}},
  \bibinfo{pages}{033641} (\bibinfo{year}{2012}).

\bibitem{CiracZollerSimulation}
\bibinfo{author}{Cirac, J.~I.} \& \bibinfo{author}{Zoller, P.}
\newblock \bibinfo{title}{Goals and opportunities in quantum simulation}.
\newblock \emph{\bibinfo{journal}{Nature Phys.}} \textbf{\bibinfo{volume}{8}},
  \bibinfo{pages}{264} (\bibinfo{year}{2012}).

\bibitem{Trust}
\bibinfo{author}{Hauke, P.}, \bibinfo{author}{Cucchietti, F.~M.},
  \bibinfo{author}{Tagliacozzo, L.}, \bibinfo{author}{Deutsch, I.} \&
  \bibinfo{author}{Lewenstein, M.}
\newblock \bibinfo{title}{Can one trust quantum simulators?}
\newblock \emph{\bibinfo{journal}{Rep. Prog. Phys.}}
  \textbf{\bibinfo{volume}{75}}, \bibinfo{pages}{082401}
  (\bibinfo{year}{2012}).

\bibitem{BlochSimulator}
\bibinfo{author}{Bloch, I.}, \bibinfo{author}{Dalibard, J.} \&
  \bibinfo{author}{Nascimbene, S.}
\newblock \bibinfo{title}{Quantum simulations with ultracold quantum gases}.
\newblock \emph{\bibinfo{journal}{Nature Phys.}} \textbf{\bibinfo{volume}{8}},
  \bibinfo{pages}{267} (\bibinfo{year}{2012}).

\bibitem{MonteCarloValidator}
\bibinfo{author}{Trotzky, S.} \emph{et~al.}
\newblock \bibinfo{title}{{Suppression of the critical temperature for
  superfluidity near the Mott transition: validating a quantum simulator}}.
\newblock \emph{\bibinfo{journal}{Nature Phys.}} \textbf{\bibinfo{volume}{6}},
  \bibinfo{pages}{998} (\bibinfo{year}{2010}).

\bibitem{Esslinger2010}
\bibinfo{author}{Esslinger, T.}
\newblock \bibinfo{title}{{Fermi-Hubbard physics with atoms in an optical
  lattice}}.
\newblock \emph{\bibinfo{journal}{Ann. Rev. Cond. Mat. Phys.}}
  \textbf{\bibinfo{volume}{1}}, \bibinfo{pages}{129} (\bibinfo{year}{2010}).

\bibitem{Koehl2014}
\bibinfo{author}{Pertot, D.} \emph{et~al.}
\newblock \bibinfo{title}{{Relaxation dynamics of a Fermi gas in an optical superlattice}}.
\newblock \bibinfo{note}{ArXiv:1407.6037}.

 
\bibitem{Naegerl_Ising}
\bibinfo{author}{Meinert, F.} \emph{et~al.}
\newblock \bibinfo{title}{Quantum quench in an atomic one-dimensional ising
  chain}.
\newblock \emph{\bibinfo{journal}{Phys. Rev. Lett.}}
  \textbf{\bibinfo{volume}{111}}, \bibinfo{pages}{053003}
  (\bibinfo{year}{2013}).

\bibitem{Gauge}
\bibinfo{author}{Zohar, E.}, \bibinfo{author}{Cirac, J.~I.} \&
  \bibinfo{author}{Reznik, B.}
\newblock \bibinfo{title}{A cold-atom quantum simulator for su(2) Yang-Mills
  lattice gauge theory}.
\newblock \emph{\bibinfo{journal}{Phys. Rev. Lett.}}
  \textbf{\bibinfo{volume}{110}}, \bibinfo{pages}{125304}
  (\bibinfo{year}{2013}).

\bibitem{Hofferberth_etal07}
\bibinfo{author}{Hofferberth, S.}, \bibinfo{author}{Lesanovsky, I.},
  \bibinfo{author}{Fischer, B.}, \bibinfo{author}{Schumm, T.} \&
  \bibinfo{author}{Schmiedmayer, J.}
\newblock \bibinfo{title}{{Non-equilibrium coherence dynamics in
  one-dimensional Bose gases}}.
\newblock \emph{\bibinfo{journal}{Nature}} \textbf{\bibinfo{volume}{449}},
  \bibinfo{pages}{324} (\bibinfo{year}{2007}).

\bibitem{Kinoshita_etal06}
\bibinfo{author}{Kinoshita, T.}, \bibinfo{author}{Wenger, T.} \&
  \bibinfo{author}{Weiss, D.~S.}
\newblock \bibinfo{title}{{A quantum Newton's cradle}}.
\newblock \emph{\bibinfo{journal}{Nature}} \textbf{\bibinfo{volume}{440}},
  \bibinfo{pages}{900} (\bibinfo{year}{2006}).

\bibitem{BlattSimulator}
\bibinfo{author}{Blatt, R.} \& \bibinfo{author}{Roos, C.~F.}
\newblock \bibinfo{title}{Quantum simulations with trapped ions}.
\newblock \emph{\bibinfo{journal}{Nature Phys.}} \textbf{\bibinfo{volume}{8}},
  \bibinfo{pages}{277} (\bibinfo{year}{2012}).

\bibitem{Islam_etal11}
\bibinfo{author}{Islam, R.} \emph{et~al.}
\newblock \bibinfo{title}{Onset of a quantum phase transition with a trapped
  ion quantum simulator}.
\newblock \emph{\bibinfo{journal}{Nat. Commun.}} \textbf{\bibinfo{volume}{2}},
  \bibinfo{pages}{377} (\bibinfo{year}{2011}).

\bibitem{PhysRevB.25.6622}
\bibinfo{author}{Maricq, M.~M.}
\newblock \bibinfo{title}{{Application of average Hamiltonian theory to the NMR
  of solids}}.
\newblock \emph{\bibinfo{journal}{Phys. Rev. B}} \textbf{\bibinfo{volume}{25}},
  \bibinfo{pages}{6622} (\bibinfo{year}{1982}).

\bibitem{Arimondo}
\bibinfo{author}{Arimondo, E.}, \bibinfo{author}{Ciampini, D.},
  \bibinfo{author}{Eckardt, A.}, \bibinfo{author}{Holthaus, M.} \&
  \bibinfo{author}{Morsch, O.}
\newblock \bibinfo{title}{Kilohertz-driven Bose-Einstein condensates in optical
  lattices}.
\newblock \emph{\bibinfo{journal}{Adv. At. Mol. Opt. Phys.}}
  \textbf{\bibinfo{volume}{61}}, \bibinfo{pages}{515} (\bibinfo{year}{2012}).

\bibitem{PeriodicDriving3}
\bibinfo{author}{Goldman, N.} \& \bibinfo{author}{Dalibard, J.}
\newblock \bibinfo{title}{{Periodically-driven quantum systems: Effective
  Hamiltonians and engineered gauge fields}}.
\newblock \bibinfo{note}{ArXiv:1404.4373}.

\bibitem{PeriodicDriving2}
\bibinfo{author}{Kitagawa, T.}, \bibinfo{author}{Berg, E.},
  \bibinfo{author}{Rudner, M.} \& \bibinfo{author}{Demler, E.}
\newblock \bibinfo{title}{Topological characterization of periodically driven
  quantum systems}.
\newblock \emph{\bibinfo{journal}{Phys. Rev. B}} \textbf{\bibinfo{volume}{82}},
  \bibinfo{pages}{235114} (\bibinfo{year}{2010}).

\bibitem{ExperimentFloquetDriving1}
\bibinfo{author}{Jotzu, G.} \emph{et~al.}
\newblock \bibinfo{title}{{Experimental realization of the topological Haldane
  model}}.
\newblock \bibinfo{note}{ArXiv:1406.7874}.

\bibitem{ExperimentFloquetDriving2}
\bibinfo{author}{Aidelsburger, M.} \emph{et~al.}
\newblock \bibinfo{title}{{Revealing the topology of Hofstadter bands with
  ultra-cold atoms}}.
\newblock \bibinfo{note}{ArXiv:1407.4205}.

\bibitem{ExperimentFloquetDriving3}
\bibinfo{author}{Struck, J.}, \bibinfo{author}{Simonet, J.} \&
  \bibinfo{author}{Sengstock, K.}
\newblock \bibinfo{title}{Spin orbit coupling in periodically driven optical
  lattices}.
\newblock \bibinfo{note}{ArXiv:1407.1953}.

\bibitem{Spielman}
\bibinfo{author}{Goldman, N.} \emph{et~al.}
\newblock \bibinfo{title}{Direct imaging of topological edge states in
  cold-atom systems}.
\newblock \emph{\bibinfo{journal}{PNAS}} \textbf{\bibinfo{volume}{110}},
  \bibinfo{pages}{6736} (\bibinfo{year}{2013}).

\bibitem{Moessner}
\bibinfo{author}{Lazarides, A.}, \bibinfo{author}{Das, A.} \&
  \bibinfo{author}{Moessner, R.}
\newblock \bibinfo{title}{Periodic thermodynamics of isolated systems}.
\newblock \emph{\bibinfo{journal}{Phys. Rev. Lett.}}
  \textbf{\bibinfo{volume}{112}}, \bibinfo{pages}{150401}
  (\bibinfo{year}{2014}).

\bibitem{1303.5652}
\bibinfo{author}{Endres, M.} \emph{et~al.}
\newblock \bibinfo{title}{Single-site- and single-atom-resolved measurement of
  correlation functions}.
\newblock \bibinfo{note}{ArXiv:1303.5652}.

\bibitem{Unruh2}
\bibinfo{author}{Martin-Martinez, E.}, \bibinfo{author}{Fuentes, I.} \&
  \bibinfo{author}{Mann, R.~B.}
\newblock \bibinfo{title}{{Using Berry's phase to detect the Unruh effect at
  lower accelerations}}.
\newblock \emph{\bibinfo{journal}{Phys. Rev. Lett.}}
  \textbf{\bibinfo{volume}{107}}, \bibinfo{pages}{131301}
  (\bibinfo{year}{2011}).

\bibitem{1402.6716}
\bibinfo{author}{Agarwal, K.} \emph{et~al.}
\newblock \bibinfo{title}{Chiral prethermalization in supersonically split
  condensates}.
\newblock \bibinfo{note}{ArXiv: 1402.6716}.

\bibitem{ColdAtomsInCavities}
\bibinfo{author}{Ritsch, H.}, \bibinfo{author}{Domokos, P.},
  \bibinfo{author}{Brennecke, F.} \& \bibinfo{author}{Esslinger, T.}
\newblock \bibinfo{title}{Cold atoms in cavity-generated dynamical optical
  potentials}.
\newblock \emph{\bibinfo{journal}{Rev. Mod. Phys.}}
  \textbf{\bibinfo{volume}{85}}, \bibinfo{pages}{553} (\bibinfo{year}{2013}).

\bibitem{SmallestEngines}
Linden, N., Popescu, S.  \& Skrzypczyk, P. 
How small can thermal machines be? The smallest possible refrigerator, \emph{Phys. Rev. Lett.} {\bf 105}, 130401 (2010).

\bibitem{Gallego}
\bibinfo{author}{Gallego, R.}, \bibinfo{author}{Riera, A.} \&
  \bibinfo{author}{Eisert, J.}
\newblock \bibinfo{title}{Correlated thermal machines in the micro-world}
  (\bibinfo{year}{2013}).
\newblock \bibinfo{note}{Arxiv.1310.8349}.

\bibitem{AdScft}
\bibinfo{author}{Hubeny, V.~E.} \& \bibinfo{author}{Rangamani, M.}
\newblock \bibinfo{title}{A holographic view on physics out of equilibrium}.
\newblock \emph{\bibinfo{journal}{Adv. High En. Phys.}}
  \textbf{\bibinfo{volume}{2010}}, \bibinfo{pages}{297916}
  (\bibinfo{year}{2010}).

\bibitem{Intensive}
\bibinfo{author}{Kliesch, M.}, \bibinfo{author}{Gogolin, C.},
  \bibinfo{author}{Kastoryano, M.~J.}, \bibinfo{author}{Riera, A.} \&
  \bibinfo{author}{Eisert, J.}
\newblock \bibinfo{title}{Locality of temperature}.
\newblock \emph{\bibinfo{journal}{Phys. Rev. X}} \textbf{\bibinfo{volume}{4}},
  \bibinfo{pages}{031019} (\bibinfo{year}{2014}).

\bibitem{DissipationZoller}
\bibinfo{author}{Diehl, S.} \emph{et~al.}
\newblock \bibinfo{title}{Quantum states and phases in driven open quantum
  systems with cold atoms}.
\newblock \emph{\bibinfo{journal}{Nature Phys.}} \textbf{\bibinfo{volume}{4}},
  \bibinfo{pages}{878} (\bibinfo{year}{2008}).

\bibitem{BlattOpen}
\bibinfo{author}{Barreiro, J.~T.} \emph{et~al.}
\newblock \bibinfo{title}{An open-system quantum simulator with trapped ions}.
\newblock \emph{\bibinfo{journal}{Nature}} \textbf{\bibinfo{volume}{470}},
  \bibinfo{pages}{486} (\bibinfo{year}{2011}).

\bibitem{OttDissipation}
\bibinfo{author}{Barontini, G.} \emph{et~al.}
\newblock \bibinfo{title}{Controlling the dynamics of an open many-body quantum
  system with localized dissipation}.
\newblock \emph{\bibinfo{journal}{Phys. Rev. Lett.}}
  \textbf{\bibinfo{volume}{110}}, \bibinfo{pages}{035302}
  (\bibinfo{year}{2013}).

\bibitem{DissipationTopologyZoller}
\bibinfo{author}{Diehl, S.}, \bibinfo{author}{Rico, E.},
  \bibinfo{author}{Baranov, M.~A.} \& \bibinfo{author}{Zoller, P.}
\newblock \bibinfo{title}{Topology by dissipation in atomic quantum wires}.
\newblock \emph{\bibinfo{journal}{Nature Phys.}} \textbf{\bibinfo{volume}{7}},
  \bibinfo{pages}{971} (\bibinfo{year}{2011}).

\end{thebibliography}
\end{document}